\documentclass[5p,times]{elsarticle}

\usepackage{lineno,hyperref}
\usepackage{gensymb}
\usepackage{amsmath}
\modulolinenumbers[5]

\journal{}

\begin{document}

\begin{frontmatter}

\title{First-principles study of CO and OH adsorption on
In-doped ZnO surfaces}

\author[mymainaddress]{R. Saniz}
\author[mymainaddress]{N. Sarmadian}
\author[mymainaddress]{B. Partoens}
\author[mysecondaryaddress]{M. Batuk}
\author[mysecondaryaddress]{J. Hadermann}
\author[mysecondaryaddressII]{A. Marikutsa}
\author[mysecondaryaddressII]{M. Rumyantseva}
\author[mysecondaryaddressII]{A. Gaskov}
\author[mysecondaryaddress]{D. Lamoen}

%\author[mymainaddress]{D. Lamoen\corref%{mycorrespondingauthor}}


\address[mymainaddress]{CMT, Departement Fysica,
Universiteit Antwerpen,
Groenenborgerlaan 171, B-2020 Antwerpen, Belgium}
\address[mysecondaryaddress]{EMAT, Departement Fysica, Universiteit
Antwerpen, Groenenborgerlaan 171, B-2020 Antwerpen, Belgium}
\address[mysecondaryaddressII]{Chemistry Department,
Moscow State University, Vorobyevy gory 1-3, Moscow 119991,
Russian Federation}

\begin{abstract}
We present a first-principles computational study of CO and OH
adsorption on non-polar
ZnO (10$\bar{1}$0) surfaces doped with indium. The calculations
were performed using a model ZnO slab. The position of the In
dopants was varied from deep bulk-like layers to the surface
layers. It was established that the preferential location of the In
atoms is at the surface by examining the dependence of the defect formation energy as well as the surface energy on In location.
The adsorption sites on the surface of ZnO and the energy of
adsorption of CO molecules and OH-species were determined in
connection to In doping. It was found that OH has higher bonding energy to the surface than CO. The presence of In atoms at the surface of ZnO is favorable for CO adsorption, resulting in an elongation of the C-O bond and in charge transfer to the surface.
The effect of CO and OH adsorption on the electronic and conduction properties of surfaces was assessed. We conclude that In-doped ZnO
surfaces should present a higher electronic response upon
adsorption of CO. 
\end{abstract}

\begin{keyword}
Zinc oxide; doping; surfaces; first-principles methods; gas sensor; carbon monoxide

\end{keyword}

\end{frontmatter}

%\linenumbers

\section{Introduction}

Zinc oxide is a $n$-type, wide-band gap (3.2 eV)
semiconductor that has long been of interest for
resistive gas sensors, as well as for a range of other
applications, such as in chemical production, catalysis, photocatalysis, pharmaceutics, etc.
\cite{shulin08,kolodziejczak14,seiyama66,ellmer10,pearton05}.
Functioning of such sensors is based on the modification of electrical properties of the semiconductor by the adsorption of gas molecules at the surface. Of particular interest are nanostructured materials with high surface-to-volume ratio, thus having enhanced gas sensitivity
\cite{pearton05,wan04,hjiri14,kaneti14}.
However, nanocrystalline pristine ZnO has poor electronic conduction because of intrinsically low charge carrier concentration and restrictions due to the dimensions of the particles \cite{ellmer10,pearton05}.
A common approach to modify its electronic conduction is
$n$-doping by group III metals, i.e., Al, Ga,
and In \cite{ellmer10,hjiri14,paraguay00,hjiri15}.
Due to the substantial discrepancy of ionic radii,
the applicability of Al(III) for substitution of Zn(II) is limited \cite{pearton05}. Indeed, the
effect of aluminum on the sensitivity of Al-doped ZnO
to CO and volatile organic compounds was observed to be minor
\cite{lim15,hsu17}. This
can be rationalized by the high solubility of the small
Al(III) cations in bulk ZnO and the hindering of their migration to the surface. Instead, doping by Ga(III) and In(III), that have a ionic radii closer to that of Zn(II), is more often in
use \cite{ellmer10,pearton05}.
It was demonstrated that doping ZnO with Ga improved not only the conductivity, but also the sensitivity to various toxic or flammable gases, e.g., CO, formaldehyde, H$_2$, ammonia, and
methane \cite{hjiri15,hou14,nulhakim17,phan13,kim09,han09}.
The influence of doping on sensitivity is often attributed to
electronic effects at the semiconductor
surface \cite{hou14,hjiri15},
while the chemical aspect of gas molecules reception on the
doped ZnO surface is overlooked.
In order to tailor the functional properties of doped
ZnO-based sensors, it is important to understand the effect of the dopants on the interaction between the surface and adsorbed target gas molecules. In a recent work, we
demonstrated correlations between the sensitivity to
H$_2$S and NO$_2$ and the
surface acidity and the paramagnetic donor sites controlled by dopant content in ZnO(Ga) \cite{vorobyeva13}. First-principles
computational modelling based on density-functional theory (DFT)
provide  
an efficient tool for the investigation of gas-solid interactions.
For instance, the effect of dopant on CO adsorption energy and preferred adsorption position was evaluated for model ZnO(Ga) clusters
\cite{derakhshandeh16}.
Of chief interest from the modelling point of view are the preferred impurity atoms position, defect formation energies, charge transfer between adsorbed molecules and sensor surface, and the change in the surface charge carrier density upon adsorption \cite{derakhshandeh16,khuili16}.
The present work aims at investigating the effect of In doping of nanocrystalline ZnO on the material structure, the interaction with CO gas and OH molecules, and the sensing behavior.
CO is one of the most common and hazardous reductive air contaminants, while OH-groups are present at the oxide surface due to chemisorption of water from ambient humidity and interfere with the sensitivity to CO target molecules.
Here we present the results of a first-principles study of a model ZnO slab
exposing the non-polar (10$\bar 1$0) surface.

Section II presents the methods used. Section III discusses
main results. Section IV presents our conclusions and
an outlook on future work.

\section{Methodology}

All calculations were performed within density functional theory (DFT) \cite{1,2},
using the plane-wave basis set and the projector augmented-wave method \cite{3,31}
as implemented in the Vienna Ab initio Simulation Package (VASP)\cite{4,5}.
We use DFT$+U$ \cite{8,9} and the Perdew-Burke-Ernzerhof (PBE) exchange and correlation potential \cite{10}.
A self-consistently calculated Hubbard $U$ parameter of 7.16 eV
is applied to the Zn atoms \cite{11}. With this value, the calculated structural properties of bulk ZnO are in good agreement with experiment.
We point out that in order to verify the reliability of our
approach, we peformed calculations based on the HSE hybrid
functional \cite{heyd04} for representative cases among the
systems studied in this work, corroborating our results
(we come back to this in Section~\ref{banda}; se also the
Appendix).
Our calculations for all systems studied, i.e., bulk, surfaces, and free molecules, include the van der Waals interaction (vdW-DFT) using the opt86b functional \cite{20,21}.

%figure1*
\begin{figure}
\begin{center}
\includegraphics*[width=0.5\hsize]{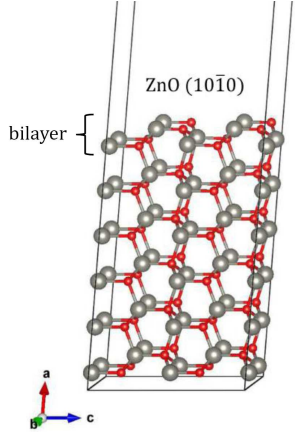}
\caption{\label{fig1} The $6\times2\times2$ supercell slab
model used for surface calculations. The slab consists of
6 bilayers. Each layer within a bilayer comprises four Zn-O
pairs. There are, thus, twelve such layers in the slab.
The vacuum region is
15 {\AA} thick.}
\end{center}
\end{figure}

We use an energy cutoff of 450 eV for the plane-wave basis set. To sample the Brillouin zone of the unit cell of bulk wurtzite
ZnO, we use a $8\times8\times4$ Monkhorst-Pack (MP) grid \cite{12} making sure that the $\Gamma$ point is included in the mesh. We note that atomic relaxations are made until residual forces on the atoms are less than 0.01 eV/{\AA} and total energies are converged to within $10^{-4}$ eV.
To study the surface
properties, we use a
$6\times 2\times 2$ supercell slab exposing
the non-polar ZnO (10$\bar{1}$0) surface.
In order to minimize
slab-slab interactions, the supercell
includes a 15 {\AA} vacuum space above this surface,
as shown in Fig.~\ref{fig1} \cite{vesta}.
The slab electronic structure is
determined using a $1\times 4\times 4$ MP grid including the
$\Gamma$ point. The surface energy $E_s$ (for either a pure or
In-doped surface) is calculated using
\begin{equation}
E_s=(E_{\rm slab}-E_{\rm supercell})/2A
\end{equation}
where $E_{\rm slab}$ and $E_{\rm supercell}$ denote the total energies of the relaxed slab (pure or In-doped) and of a bulk supercell (pure or In-doped) with an equivalent number of  formula units. $A$ is the surface area of the slab.
The slab consists
of six bilayers (see Fig.~\ref{fig1}), each consisting of
two neutral Zn-O layers (this is, twelve layers in all).
It was previously shown that with this number of layers the
surface energy is converged with respect to slab
thickness \cite{marana08}. As seen in Fig.~\ref{fig1},
each of these layers contains four Zn-O pairs.
When relaxing the slab structure, we
keep the three lowest bilayers fixed to simulate the rigidity
of the bulk-like
deeper subsurface layers in a real sample. We henceforth refer
to the upper six Zn-O layers (corresponding to the upper three bilayers) as the
six surface-like layers \cite{supplementary}. 

%figure 2
\begin{figure}
\begin{center}
\includegraphics*[width=0.8\hsize]{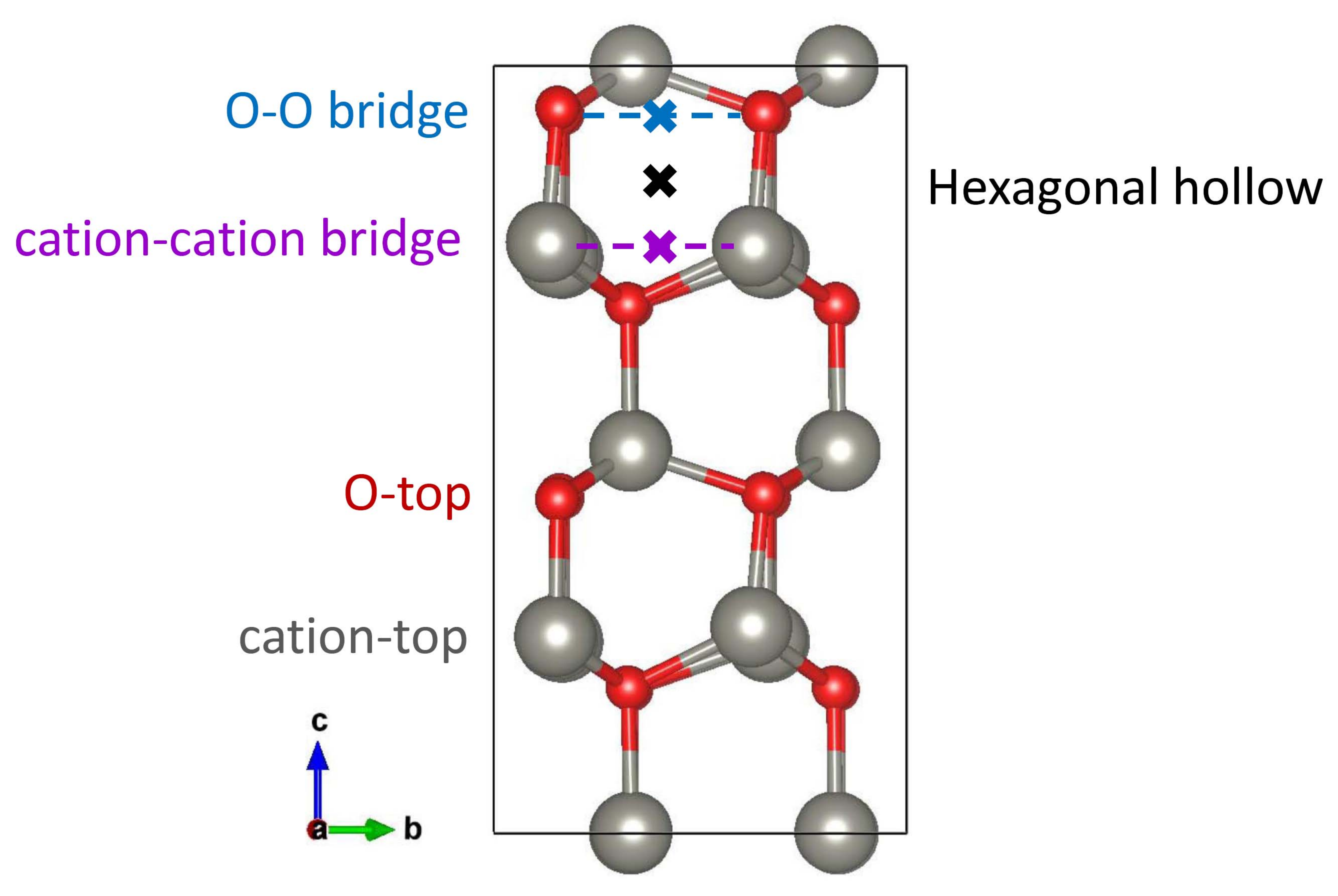}
\caption{\label{fig2} Top view of the slab supercell,
with the possible initial adsorption locations considered
here indicated (the lattice vectors in this figure are the
same as in Fig.~\ref{fig1}).}
\end{center}
\end{figure}

Regarding doping, our experimental evidence indicates that a
4-5 at.\% In concentration [i.e., [In]/([In]+[Zn])]
is sufficient for it to overcome its
solubility in ZnO and to segregate to the
surface \cite{vorobyeva17}.
Here we substitute
two Zn atoms with In in our supercell slab model,
which represents an In concentration of 4.17 at.\%.
The impurity, or dopant,
formation energy (either in the bulk or slab systems)
is given by \cite{freysoldt14}
\begin{equation}\label{fomula1}
E_{f}=E_{\rm doped}-E_{\rm pure}-n\mu_{\rm In}+n\mu_{\rm Zn},
\end{equation}
where $E_{\rm doped}$ is the total energy of the system containing
the dopants and $E_{\rm pure}$ is the total energy of pure
system (undoped slab or bulk ZnO).
$\mu_{\rm In}$ and $\mu_{\rm Zn}$ are the chemical potentials of In and Zn, respectively, and $n$ is the number of Zn atoms
substituted. The chemical potentials are calculated in their
ground state metallic phases, i.e., tetragonal for In and
hexagonal close-packed for ZnO.
When studying the effects of the location of the
dopants with respect to the topmost surface, the two In atoms
substitute two Zn atoms in a single surface-like
layer at a time. The two substitution sites in a layer are
determined by minimizing the total energy.
We denote $S_i$ ($i=1,\ldots, 6$)
the slab with the In
atoms sitting in its $i$-th surface-like layer ($S_1$
being the case where the In atoms are in
the topmost layer).

In order to study the adsorption of CO or OH, we place
one such a molecule on top of the relaxed ZnO slab, at a
distance of 2 {\AA} from the surface. We consider six possible
initial positions, namely
cation-top (in the case of a doped surface the cation can
be Zn or In),
O-top, center of hexagonal hollow,
cation-cation-bridge, and O-O-bridge. The possible
adsorption sites are shown in Fig.~\ref{fig2}.
The positions of the CO/OH molecule and of atoms in
the surface-like layers of the slab are optimized. The 
adsorption energy $E_b$ of the adsorbed molecule is defined as
\begin{equation}
E_a = E_{\rm CO/OH-ZnO}-E_{\rm CO/OH}-E_{\rm ZnO},
\end{equation}
where $E_{\rm CO/OH-ZnO}$ denotes
the total energy of the slab with the adsorbed molecule,
and
$E_{\rm CO/OH}$ and $E_{\rm ZnO}$ denote the total
energies of the free CO/OH molecule and of the ZnO slab, respectively.
Note that
the adsorption of one CO/OH molecule on top of a ($2\times2$) surface is equivalent to $1/4$ monolayer (ML) molecular coverage. 

In addition,
in order to understand the adsorption process and the contribution of different atoms in this process, we analyze
the atomic species character of the energy bands (levels)
close to the Fermi level and band edges of the
ZnO surfaces, with and without CO/OH molecule adsorption.
Finally,
the calculated electron charge density of the optimized structure is used to calculate the electronic charge partitioned for each atom using a grid-based Bader charge analysis \cite{13,14}.
Charge transfer between an absorbed molecule and
a surface can be determined using
\begin{equation}
\Delta q = -e\Delta n=-e(n_{\rm atom}-n_{\rm valence}), \label{tr}
\end{equation}
where $n_{\rm atom}$ is the calculated number of electrons around an atom in the system studied and $n_{\rm valence}$ is the number of valence electrons considered for the calculations for the corresponding atom.
This is used to further characterize the adsorption process.

\section{Results}

%figure3*
\begin{figure}
\includegraphics*[width=0.495\hsize]{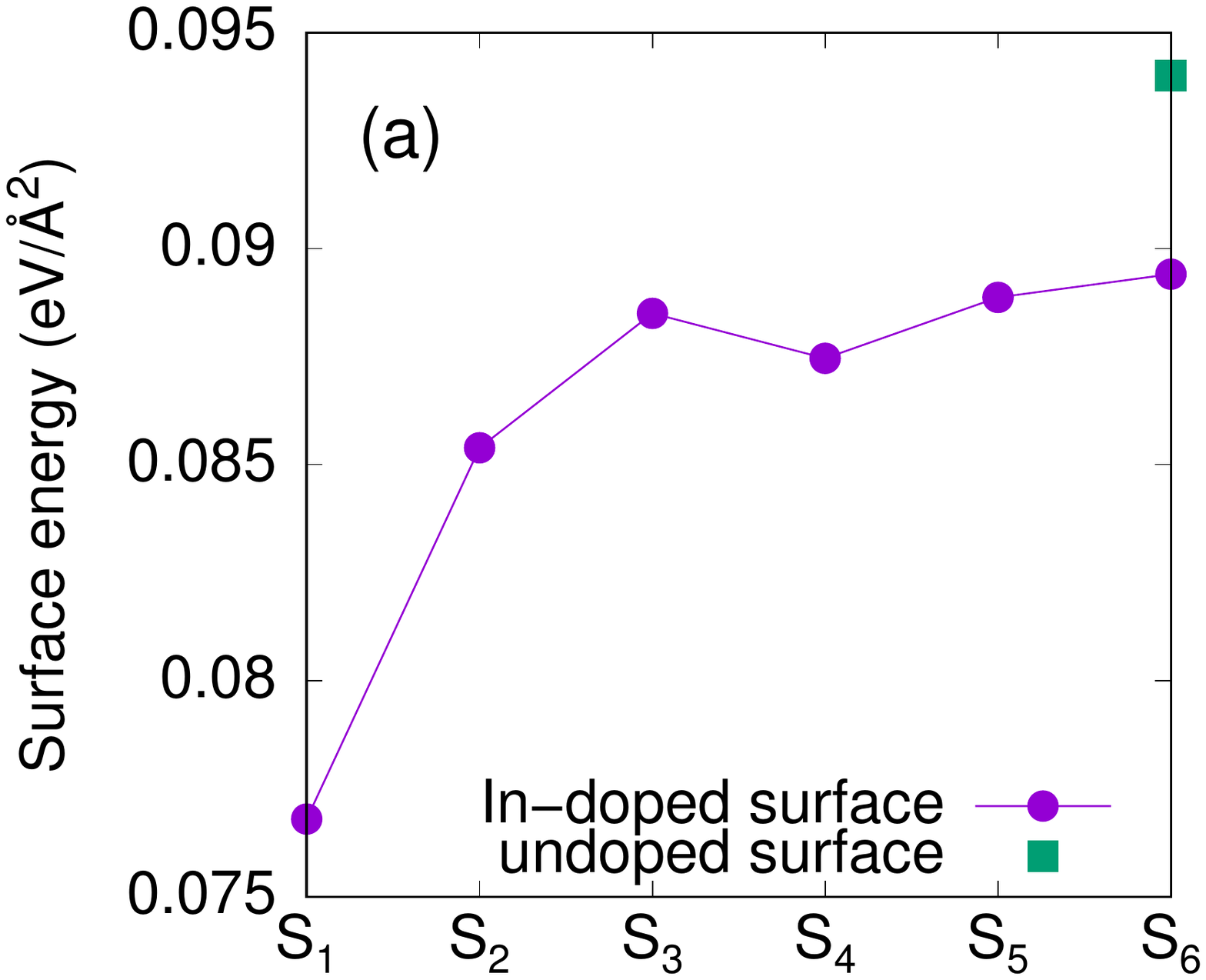}
\includegraphics*[width=0.495\hsize]{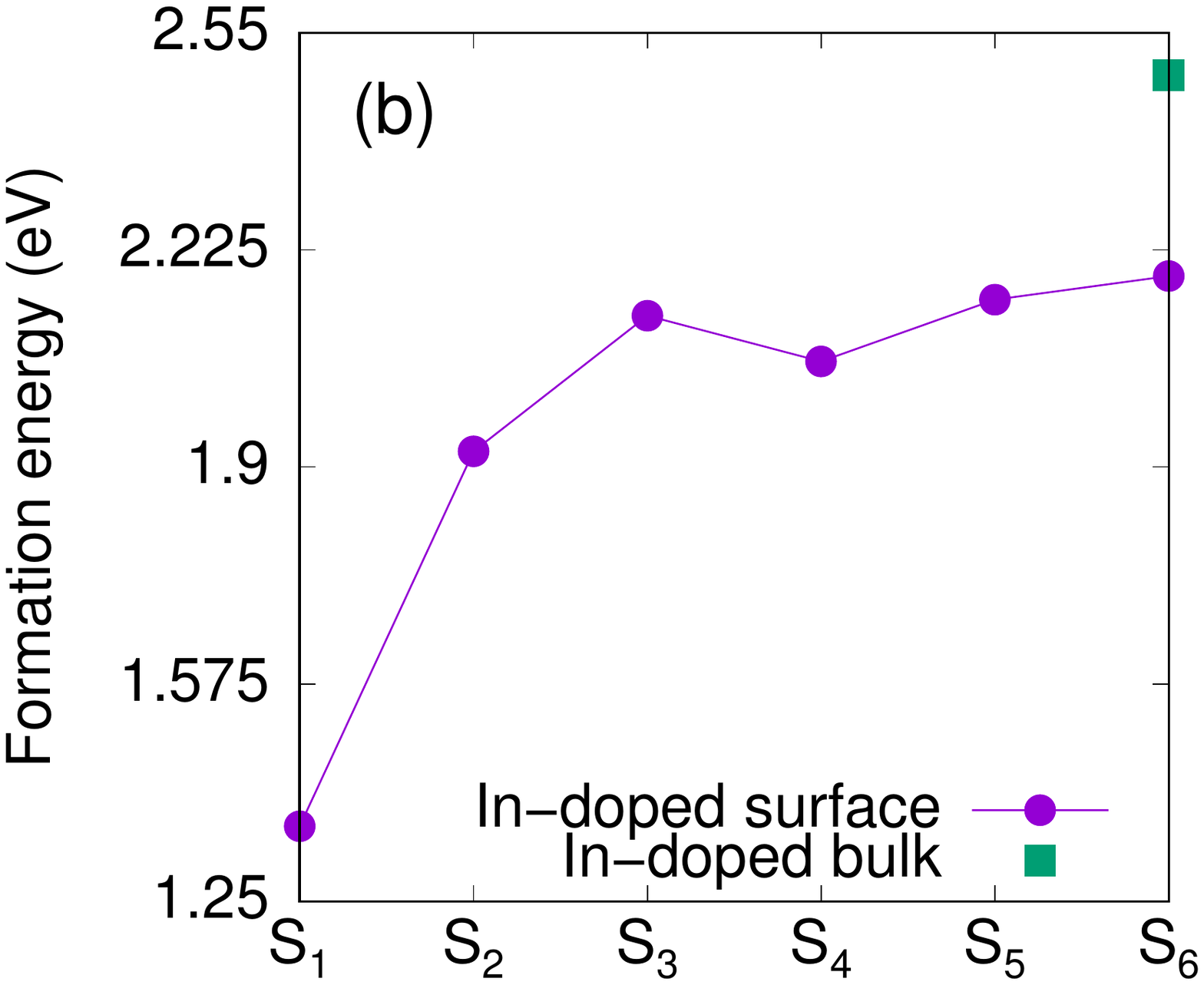}
\caption{\label{fig3} (a) Surface energy of the In-doped ZnO
$S_i$ slabs,
in comparison with the surface energy of
the undoped ZnO surface. (b) Formation energy of the
In impurities in the $S_i$ slabs, compared with the
formation energy of bulk In-doped ZnO.}
\end{figure}

\subsection{In-doped ZnO surfaces}

%figure4*
\begin{figure}
\includegraphics*[width=0.495\hsize]{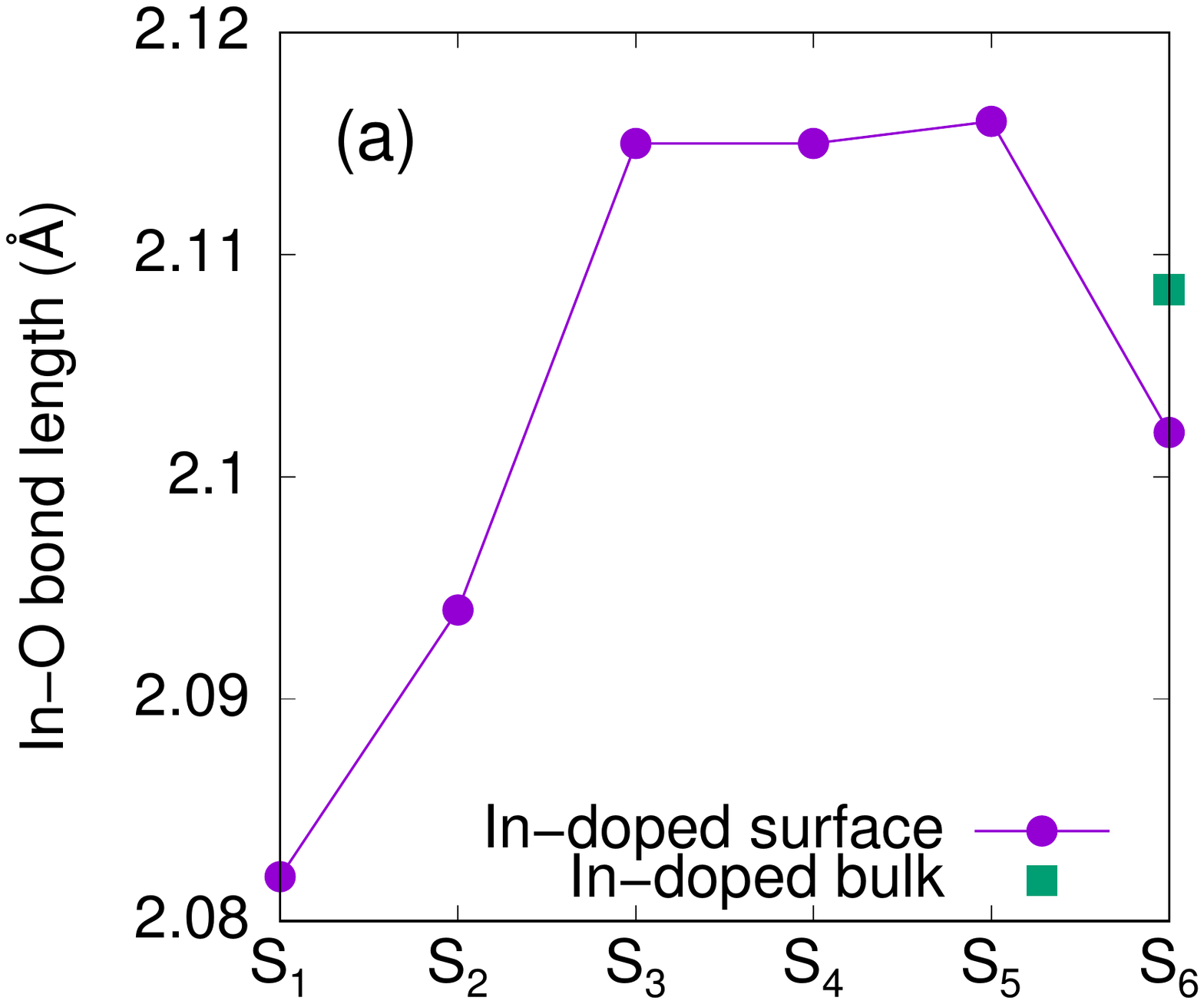}
\includegraphics*[width=0.495\hsize]{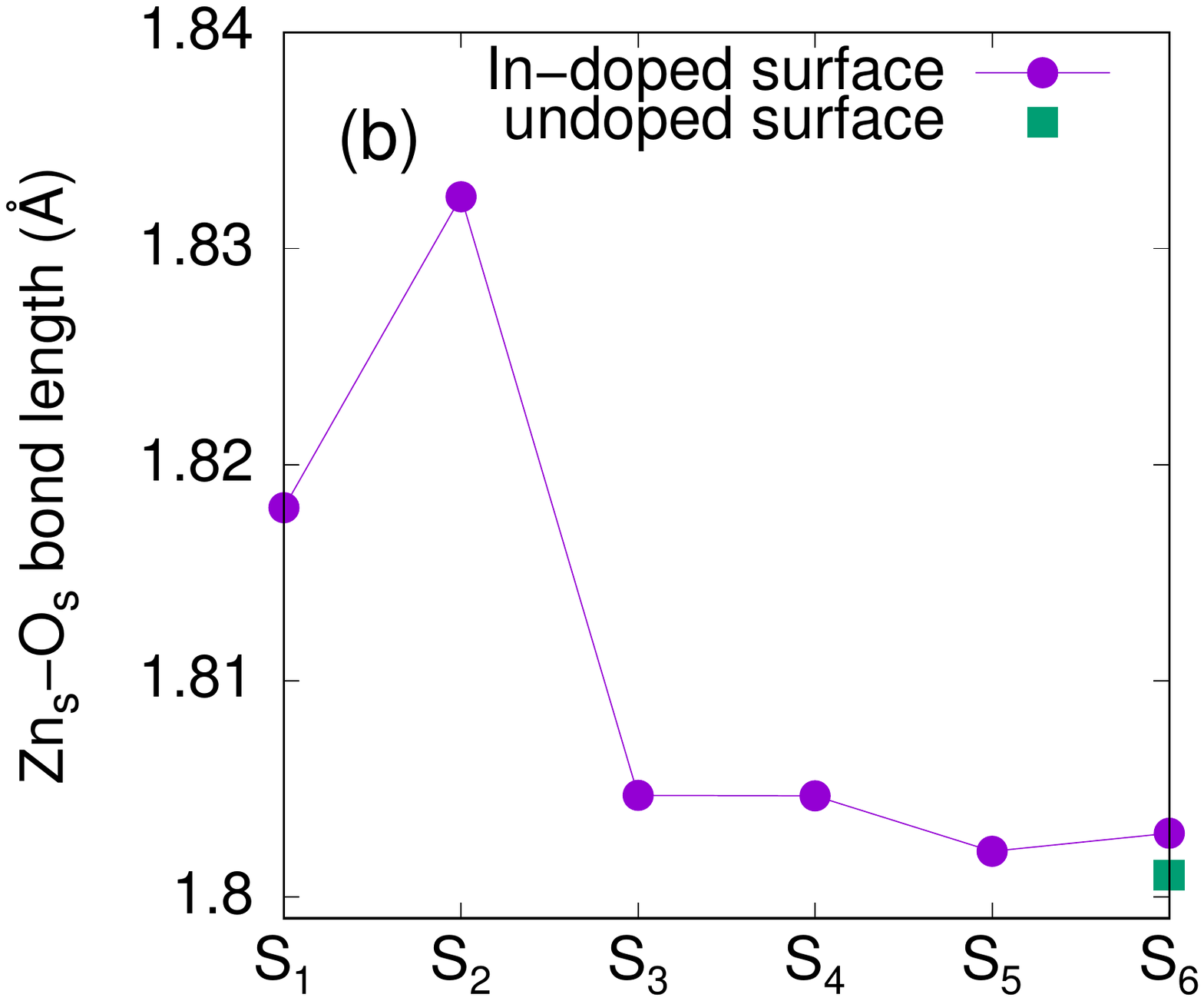}
\caption{\label{fig4} Structural change of the ZnO surfaces
with In doping. (a) Length (average) of the In-O bond and (b)
of the surface Zn to surface O bonds (Zn$_s$-O$_s$). The
corresponding bulk bond lengths are shown for comparison in
both plots.}
\end{figure}

We consider first the effect of doping on the surface
energy. For the undoped surface
we find an energy of 0.094 eV/{\AA}$^2$. This can be compared
with the value of 0.081 eV/{\AA}$^2$ obtained in
Ref.~\citenum{marana08} using the B3LYP hybrid-functional.
In Fig.~\ref{fig3}(a) we plot the surface energies of
the In-doped slabs as a function of location of the dopants,
comparing them with the energy of the undoped surface. 
As expected from the observation of In segregation
to the surface in experiment \cite{marikutsa18}, the
surface energy tends to decrease as the In dopants approach
the slab surface, reaching its lowest value
(0.077eV/{\AA}$^2$) at $S_1$.
The trend is mirrored by the dopant formation
energies, shown in Fig.~\ref{fig3}(b), where $E_f$ per
In atom is plotted. This shows more directly that the In
impurities will tend to segregate to the surface.

%figure5*
\begin{figure}
\begin{center}
\includegraphics*[width=0.9\hsize]{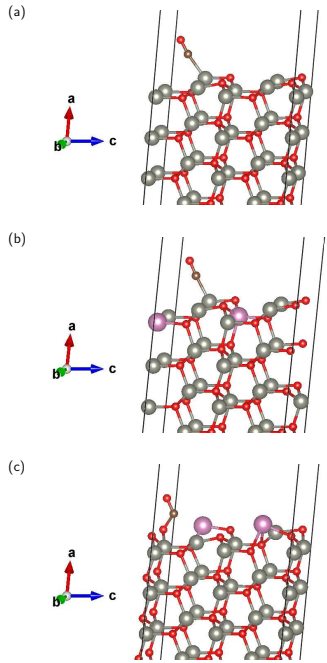}
\end{center}
\caption{\label{fig5} Structure of the relaxed CO molecule-ZnO surface systems, for the cases of (a) the undoped surface,
(b) the $S_2$ slab, and (c) the $S_1$ slab. The carbon and
indium atoms are represented by the brown and magenta
spheres, respectively. In the first two
cases CO is essentially physisorbed, with the depicted bond
being very weak (see text). In (c) the CO molecule is
chemically bond to a surface oxygen, away from the In
dopants.}
\end{figure}

We also study the surface structural changes induced
by doping. Figure~\ref{fig4}(a) shows the average
In-O bond length in the $S_i$ slabs (as noted above, there
are two In-O pairs in each case),
compared to the corresponding
bond length in a In-doped bulk supercell. Further,
we consider the average bond length of
the Zn-O pairs at the {\it surface} of the slab, which
we denote Zn$_s$-O$_s$. In Fig.~\ref{fig4}(b) we show
how that bond length depends on how far from the surface
the In dopants lie, and compares it with the same bond
length in the undoped surface. Figures~\ref{fig4}
indicate that the structural changes are more pronounced
when the In dopants lie at the surface or subsurface
of the slabs. Since these cases also present the lowest
formation energies (Figs.~\ref{fig3}), we focus on these
cases in our study below on molecule adsorption.
%{\bf We also point out that the fact that the Zn$_s$-O$_s$ bond
%lengths are elongated compared to the undoped surfaces
%can be correlated with the experimentally observed increase in
%lattice parameters in the grains, and can be understood as a
%negative chemical pressure effect (In is much larger than Zn)}.

\subsection{CO/OH adsorption}

We study the adsorption of CO and OH on undoped surfaces
as well as on the In-doped surfaces $S_1$ and $S_2$,
as indicated above. In each case we determined the preferred
adsorption site via the comparison of the total energies
corresponding to the different sites indicated in the
previous Section (see Fig.~\ref{fig2}).
Here we look
more closely into the characteristics of the preferred
sites. 

We consider first CO adsorption. The relaxed structures in the
three cases above indicated are shown in Fig.~\ref{fig5}.
In the case of the undoped surface [Fig.~\ref{fig5}(a)],
the CO molecule relaxes
to the center of the hexagonal hollow.
Note that because the CO molecule
does not stand vertically above the surface,
the C atom is closer to one of the Zn atoms forming the
hexagonal hollow, and appears bonded to it. In the
$S_2$ case the CO molecule relaxes more clearly to a
Zn-top position, and bonds with the Zn atom (the bond
length is smaller than in the previous case; see further
down). In the $S_1$ case, the CO molecule relaxes to an
O-top positions. More exactly, it relaxes to what we call
a O$_{\rm Zn}$ position. Indeed,
in the $S_1$ case, two different
O-top adsorption positions are possible: one is above of the
O atom in the In-O pair (O$_{\rm In}$), and the other
above the O atom in the Zn-O pair (O$_{\rm Zn}$).
The relaxed surface structure in the $S_1$ case is shown in
Fig.~\ref{fig5}(c). Note that the possible reaction of CO on
metal oxides surfaces
with pre-adsorbed or lattice oxygen has been 
studied many times in the past \cite{barsan01}.
Such a mechanism has been invoked, for instance, in the study of the CO sensing properties of Ti-doped SnO surfaces
\cite{zeng14}.
Furthermore, the possible reaction of CO with adsorbed oxygen has been hypothesized by the group of G. Neri in a series of experimental articles studying the CO sensing properties of ZnO doped with Al, Ga, and In \cite{hjiri14,hjiri15,dhahri17}.
Interestingly, here we find that CO can interact with a lattice
oxygen when the surface is In-doped.

%figure6*
\begin{figure}
\begin{center}
\includegraphics*[width=0.9\hsize]{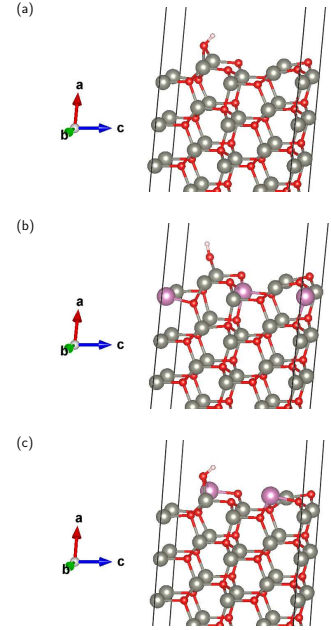}
\end{center}
\caption{\label{fig6} Structure of the relaxed OH molecule-ZnO surface systems, for the cases of (a) the undoped surface,
(b) the $S_2$ slab, and (c) the $S_1$ slab.
The hydrogen atom is represented by a white sphere (the 
indium atoms are represented by magenta spheres, as in
Figs.~\ref{fig5}). In all cases,
the OH molecule is chemically bonded to the surface via a
bridge bond (Zn-Zn bridge in (a) and (b), and Zn-In bridge
in (c).}
\end{figure}

In the case of OH adsorption,
the results are markedly
different. On the undoped surface and
on $S_2$
the OH molecule relaxes to the Zn-Zn bridge position,
although the OH molecule orientation with respect to the
surface is different. This is shown in
Figs.~\ref{fig6}(a) and (b), respectively.
In Fig.~\ref{fig6}(b) the bridge cannot be seen because
the OH molecule bridges Zn atoms in neighboring cells
along the $b$ direction, and
only the atoms in a single cell are shown.
In the case of $S_1$, the OH molecule relaxes to a
Zn-In bridge position, as seen in Fig.~\ref{fig6}(c).

\begin{figure}
\begin{center}
\includegraphics*[width=0.9\hsize]{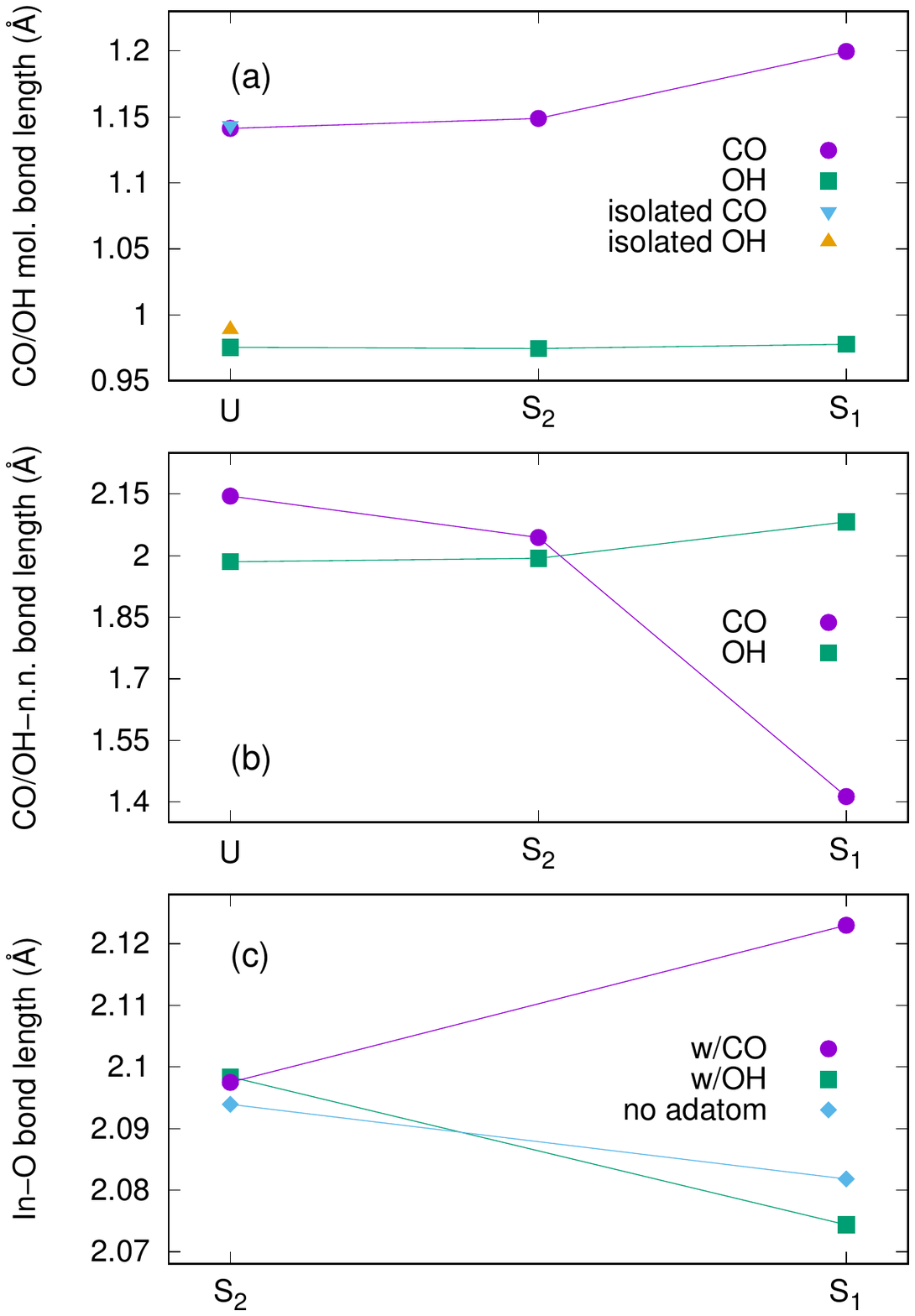}
\end{center}
\caption{\label{fig7} (a) Intramolecular bond lengths
for the CO and OH molecules adsorbed on the undoped
and $S_1$ and $S_2$ slabs. (b) Bond length of the CO or
OH molecules to their nearest neighbor in the slabs.
(c) Average In-O bond lengths in the $S_1$ and $S_2$ slabs,
compared to the bond length in the case with no adsorbed
molecule.}
\end{figure}

We now look in more detail into how In doping affects
the surface structure and how it correlates with the
adsorption energy of the adsorbed molecules and with charge
transfer between the latter and the surfaces.
In Fig.~\ref{fig7}(a) we consider the interatomic
bond length of the CO and OH molecules themselves, where
we compare the bond length of the isolated molecules
with their bond length when adsorbed on an undoped surface
(U), and on doped surfaces $S_1$ and $S_2$.
We can see that in the case of the OH molecule, the
bond length is only weakly affected by the surface,
whether the latter
is In-doped or not. In the case of the CO molecule, its
bond legth is also weakly affected in the undoped surface
and $S_2$ cases. However, in the $S_1$ case, the bond
length is elongated by as much as 5\%.

%figure7*

To complement this
information, we consider the distance between the adsorbed
molecules and the surface, i.e., the bond length between
C (O) in the CO (OH) molecule and its
nearest neighbor on the ZnO surface (in the case of the
bridge positions, we consider the average of the bond
lengths involved). These bond lengths are shown in
Fig.~\ref{fig7}(b). Again, doping affects the CO molecule
considerably more than the OH molecule. In the latter case
the average bridge bond length hardly changes between the
undoped surface and $S_2$, elongating by close to 4.9\% on
$S_1$. On the other hand, in the case of the CO molecule
the bond length is 4.7\% shorter on $S_2$ than on the
undoped surface. But on $S_1$ the bond length is
dramatically shorter, falling by 34\% with respect to the
bond length on the undoped surface. The main
reason for such a strong difference is of course that on
$S_1$ the CO molecule is bonded to a surface oxygen,
while on the undoped surface (and $S_2$) it is bonded to
zinc. Finally, it is interesting to see the effect of the
adsorbed molecules on the In-O bond lengths in the case of
the doped surfaces. This is illustrated in Fig.~\ref{fig7}(c).
We can see that the adsorbed molecules have relatively little
effect on the In-O bond length when In doping occurs in the
subsurface ($S_2$). On the other hand, in the case of $S_1$
the adsorption of CO and OH has opposite effects. The latter
decreases the bond length by less than 0.4\%, while the former
elongates it by nearly 2\%.

%figure8*
\begin{figure}
\begin{center}
\includegraphics*[width=0.9\hsize]{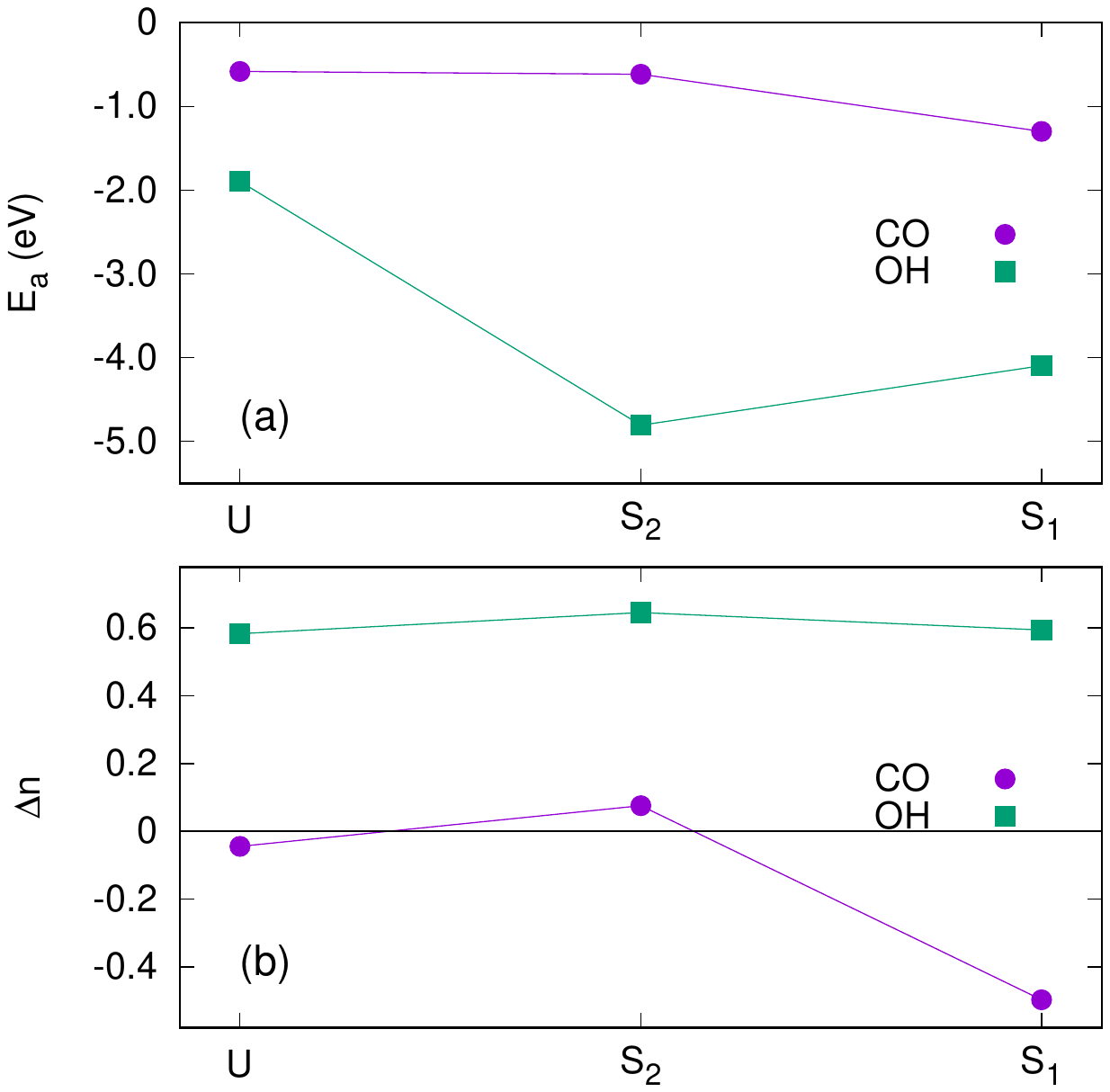}
\end{center}
\caption{\label{fig8} (a) CO and OH adsorption energies
to the undoped and $S_1$ and $S_2$ slabs.
(b) Molecule number of electrons gain or loss
for the CO and OH
molecules in the three cases indicated in (a).}
\end{figure}

In Fig.~\ref{fig8}(a) we show the adsorption energies of the
CO and OH molecules on the undoped and doped surfaces.
The value we obtain for CO adsorbed on the undoped surface
is 0.59 eV. This is a comparatively weak value, indicative of
physisorption.
This is in line with experiment, where
ultraviolet photoelectron spectroscopy measurements on
ZnO powders yield an adsorption energy value of $\sim 0.52$ eV
(note that in such experiment, polar as well as
non-polar surfaces are probed) \cite{gay80}. The value 
increases only slightly in the case of the $S_2$ surface.
In the case of the $S_1$ surface, however, the adsorption 
energy increases to 1.30 eV, indicating that CO is actually
chemically bound to the surface.
The OH adsorption energies are much stronger in all cases.
Our calculated value for the undoped surface, 1.90 eV,
compares well with previously reported values,
1.72 or 1.78 eV \cite{vines13}. As for CO, doping the
surfaces with In increases the adsorption energies.

The charge transfer between CO and OH and the
ZnO surfaces yields additional information on the nature
of the adsorption. Fig.~\ref{fig8}(b) shows that there
is strong charge transfer toward the OH molecule
[see Eq.~(\ref{tr})] in all cases, consistent with a chemical
bond. In contrast, the CO molecule shows only very weak charge
transfer in the cases of the undoped surface and surface
$S_2$, which is typical of physisorption. 
However, in the case of surface $S_1$ there is a much stronger
charge transfer, toward the surface in this case, which is
congruent with the stronger adsorption energy indicated above.

%figure9*
\begin{figure}
\begin{center}
\includegraphics*[width=0.8\hsize]{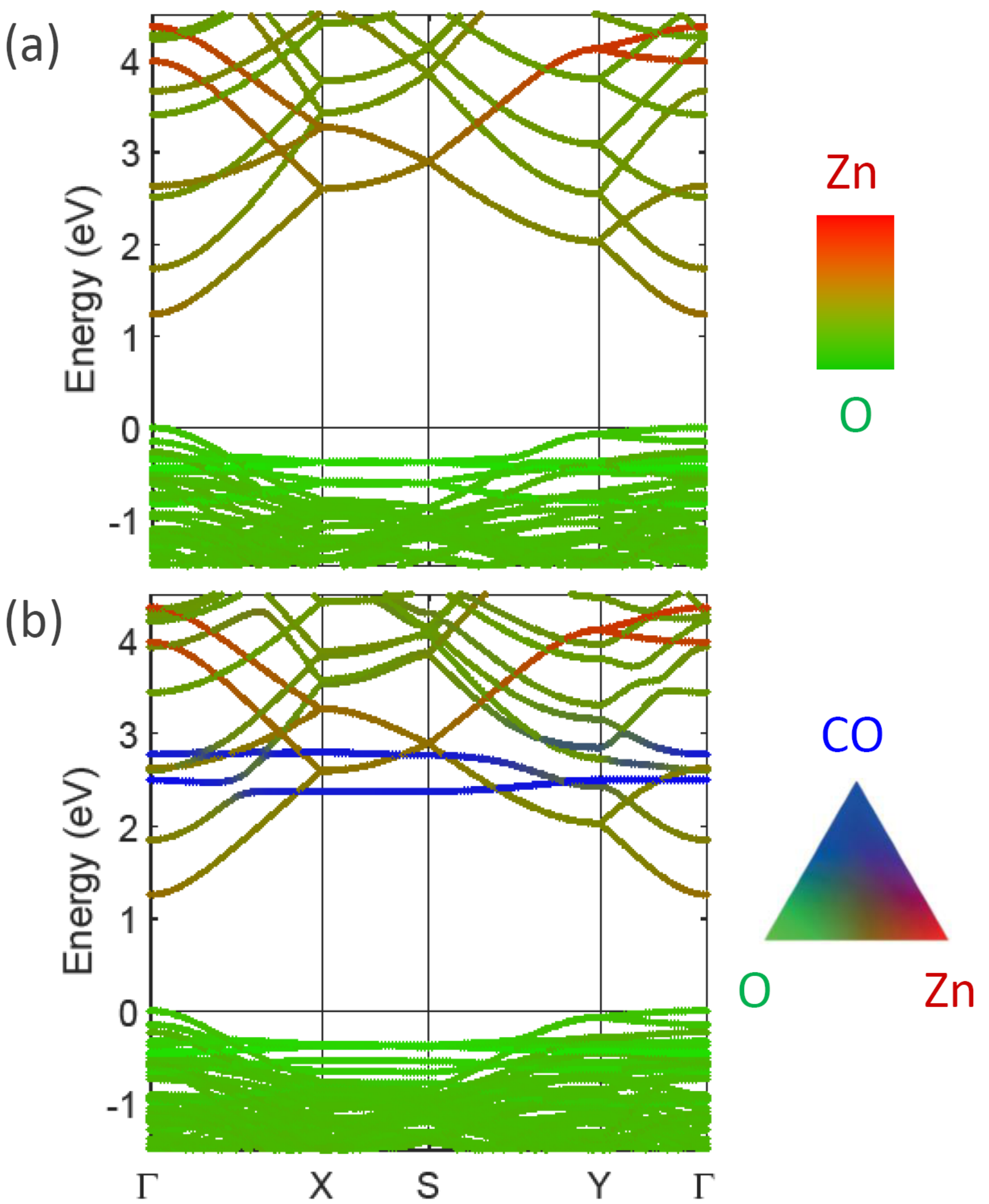}
\end{center}
\caption{\label{fig9} Character of the band structures
of (a) the undoped ZnO slab, with the colors highlighting the
Zn and O contributions (red and green, respectively),
and (b) the undoped slab with CO
adsorbed, the contribution of the latter highlighted
in blue.}
\end{figure}

\section{Effect of adsorption on surface electronic properties
and semiconductor behavior}\label{banda}

We discuss here the implications that the adsorption of gas
molecules has on the electronic properties of
In-doped ZnO surfaces.
For this we look into the character
of the states near the valence band maximum (VBM) and
conduction band minimum (CBM) of the ZnO slabs and how these
are modified by the In-doping and by the adsorption of CO and
OH.

In Fig.~\ref{fig9}(a) we plot the band structure of a pure
ZnO slab, i.e., with no In doping and with no adsorbed
molecule \cite{matlab}.
The slabs present rectangular symmetry on the
$bc$-plane, and the bands correspond to states along the
high symmetry lines on the same plane. The zero of energy
is set at the VBM. The ZnO slab remains
a semiconductor, with the VBM and CBM both at the $\Gamma$
point, as in bulk ZnO.
The color indicates the character of the bands. Thus, the
valence bands have a clearly dominant oxygen character,
while the conduction bands have more mixed character, with
some bands showing stronger zinc or oxygen contribution.
In Fig.~\ref{fig9}(b) we consider the effect of CO adsorption.
The CO bonding molecular levels are deep, well below
the minimum energy in the plot, with the antibonding levels
well above the CBM and clearly very weakly hybridized with the
ZnO levels. Thus, the valence and conduction bands are largely
left untouched. This indicates that the CO molecule is not
chemically bound to the surface, in agreement with our
assessment based on bond lengths, adsorption energy, and
charge transfer presented in the previous Section. Quite
importantly, this means that the electronic properties of pure
ZnO (10$\bar{1}$0) surfaces would be practically
unaffected by adsorbed CO
molecules and, hence, that inferior CO sensitivity can be expected from undoped ZnO surfaces.

%figure10*
\begin{figure}
\begin{center}
\includegraphics*[width=0.8\hsize]{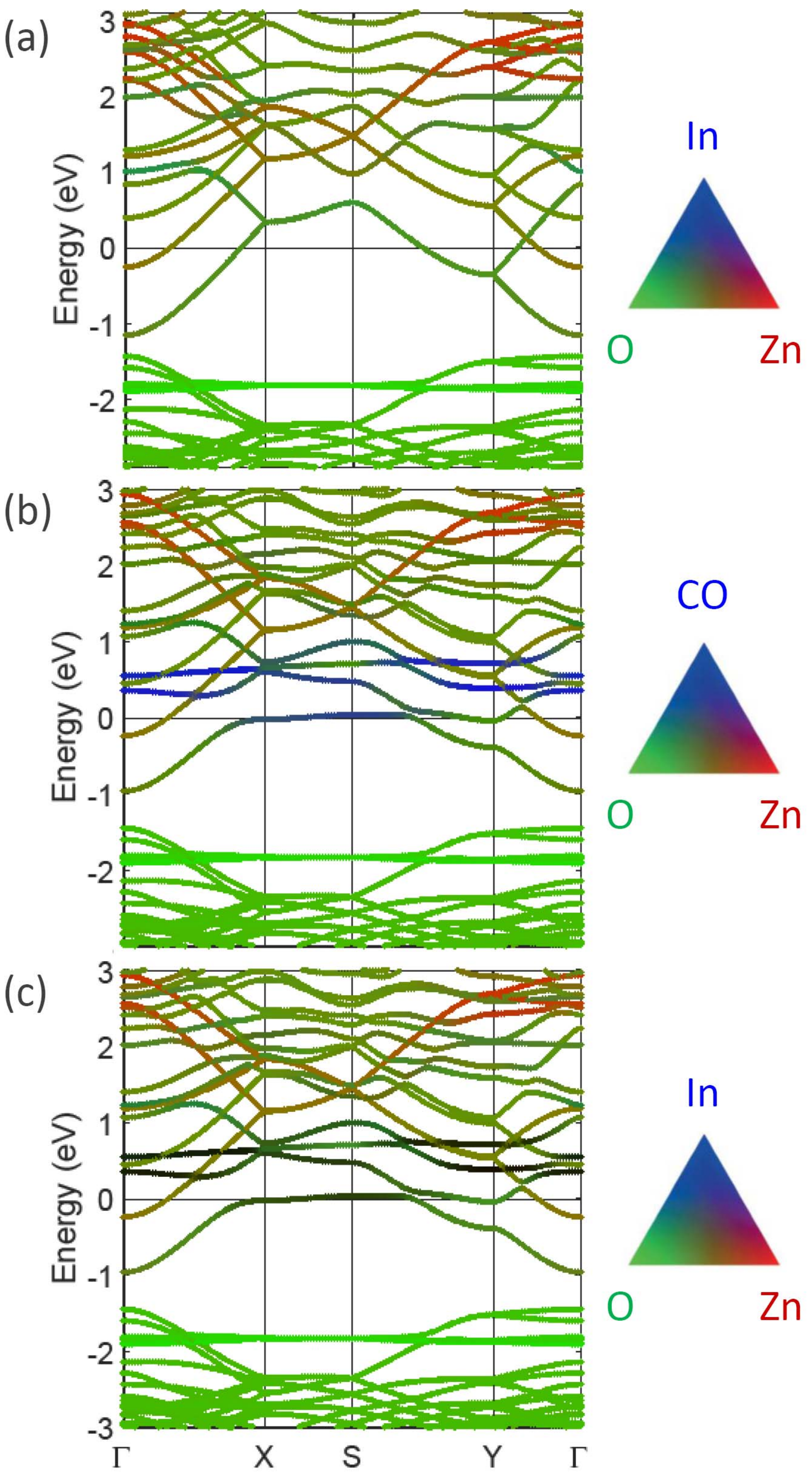}
\end{center}
\caption{\label{fig10} Character of the band structures
of (a) the In-doped ZnO slab with no CO adsorbed, with the colors highlighting the
In, Zn, and O contributions (blue, red, and green, respectively);
(b) the $S_2$ slab with CO
adsorbed, with the colors highlighting the
CO, Zn, and O contributions (blue, red, and green, respectively); (c) same as in (b), but with blue highlighting
the In contribution.}
\end{figure}

In Fig.~\ref{fig10} we consider the $S_2$ slabs.
Figure~\ref{fig10}(a) presents the band structure of a
$S_2$ slab with no CO adsorbed. This band structure is
very close to the band structure of bulk In-doped ZnO,
although the band gap narrowing here is stronger
(see, e.g., Ref.~\citenum{saniz13}). The excess In electrons
(with respect to Zn) go on to populate the lowest conduction
band, thus loosing their In character and becoming
delocalized. The $S_2$ slab, therefore,
remains an $n$-type doped semiconductor, showing no great
consequential surface effects from the electronic point of
view. Figures~\ref{fig10}(b) and (c) present the band
structure when a CO molecule is adsorbed on the surfaces
of $S_2$, highlighting (in addition to O and Zn) the contribution of the CO molecule [(b)] and of In [(c)].
Again the CO molecule states shows little hybridization with
the surface atoms. This agrees with the weak adsorption energy
and low charge transfer shown in Figs.~\ref{fig8}(a) and (b)
[note that the bond lengths in Figs.~\ref{fig7}(a) and (b)
show little change with respect to the CO molecule
adsorbed on the undoped ZnO surface; compare also
Figs.~\ref{fig5}(a) and (b)]. We can see in
Fig.~\ref{fig10}(b) that there are some CO states near the
Fermi level along the X-S high symmetry line
(the states close to S actually fall just above the Fermi
level and are empty). This gives rise to the weak positive
charge transfer in Fig.~\ref{fig8}(b) in this case.
The effect of these states on the electronic properties of the
slab, however, are expected to be weak because of the small
fraction of charge involved. Thus, we conclude that as long
as the In dopants reside at the interior of the slab, they
will not result in a surface electronically sensitive to CO.

%figure11*
\begin{figure}
\begin{center}
\includegraphics*[width=0.8\hsize]{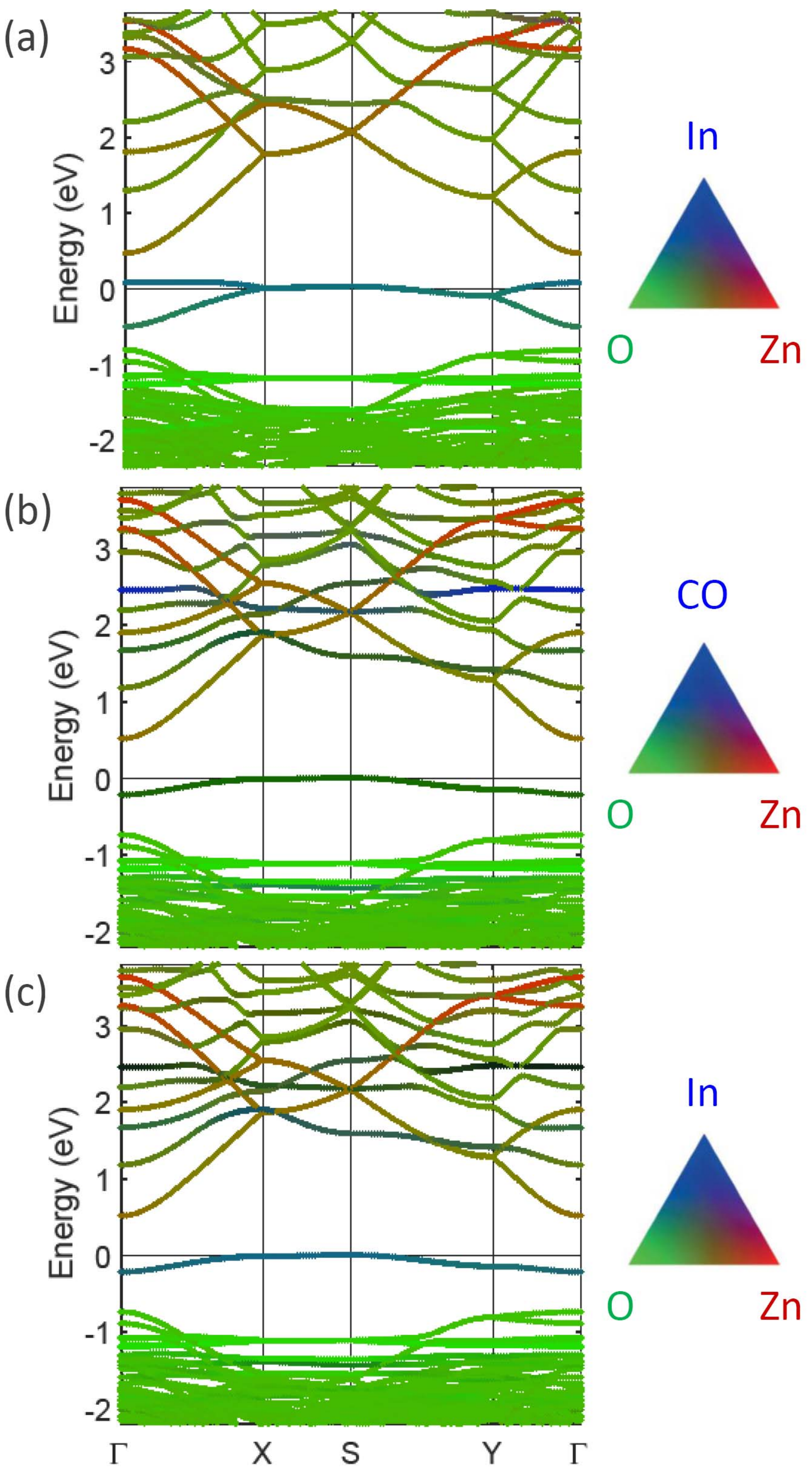}
\end{center}
\caption{\label{fig11} Same as Fig.~\ref{fig10}, but
for the case of the $S_1$ slab.}
\end{figure}

Matters change significantly in the case of the $S_1$ slab.
Figure~\ref{fig11}(a) shows the band structure of the $S_1$
slab with no CO adsorbed. It clearly indicates that the In
dopants create highly localized states in the band gap,
arising from the In dangling bonds at the surface (the
non-zero dispersion in these state in the plot is only due
to the small size of the supercell, especially in the $b$
direction). Thus, because of band bending at the surface, these
states will act as electrons traps and hinder the 
$n$-type conductivity arising from any In dopants in the
interior of the slab (case above). The effect of CO adsorption
is significant. As Figs.~\ref{fig11}(b) and (c) show, it
results essentially in passivation of the In dangling bonds
(adsorption of only one CO molecule is considered in this study), consequently
eliminating the electron traps associated to them and 
thus increasing $n$-type conductivity of the slab.
This picture is supported by a charge density
difference plot showing
how charge is redistributed between the CO molecule and the
surface, as shown in Fig.~\ref{fig14} in the Appendix.
Thus, the presence of In atoms at the surface of ZnO(In) is favorable for the electronic signal generation upon CO adsorption. The processes of adsorption of target gas molecules and of oxidation of adsorbates at the sensor surface determine the sensing behavior. Hence, the sensitivity of In-doped ZnO to CO should be improved once the dopant segregates to the surface. 

%figure12*
\begin{figure}
\begin{center}
\includegraphics*[width=0.8\hsize]{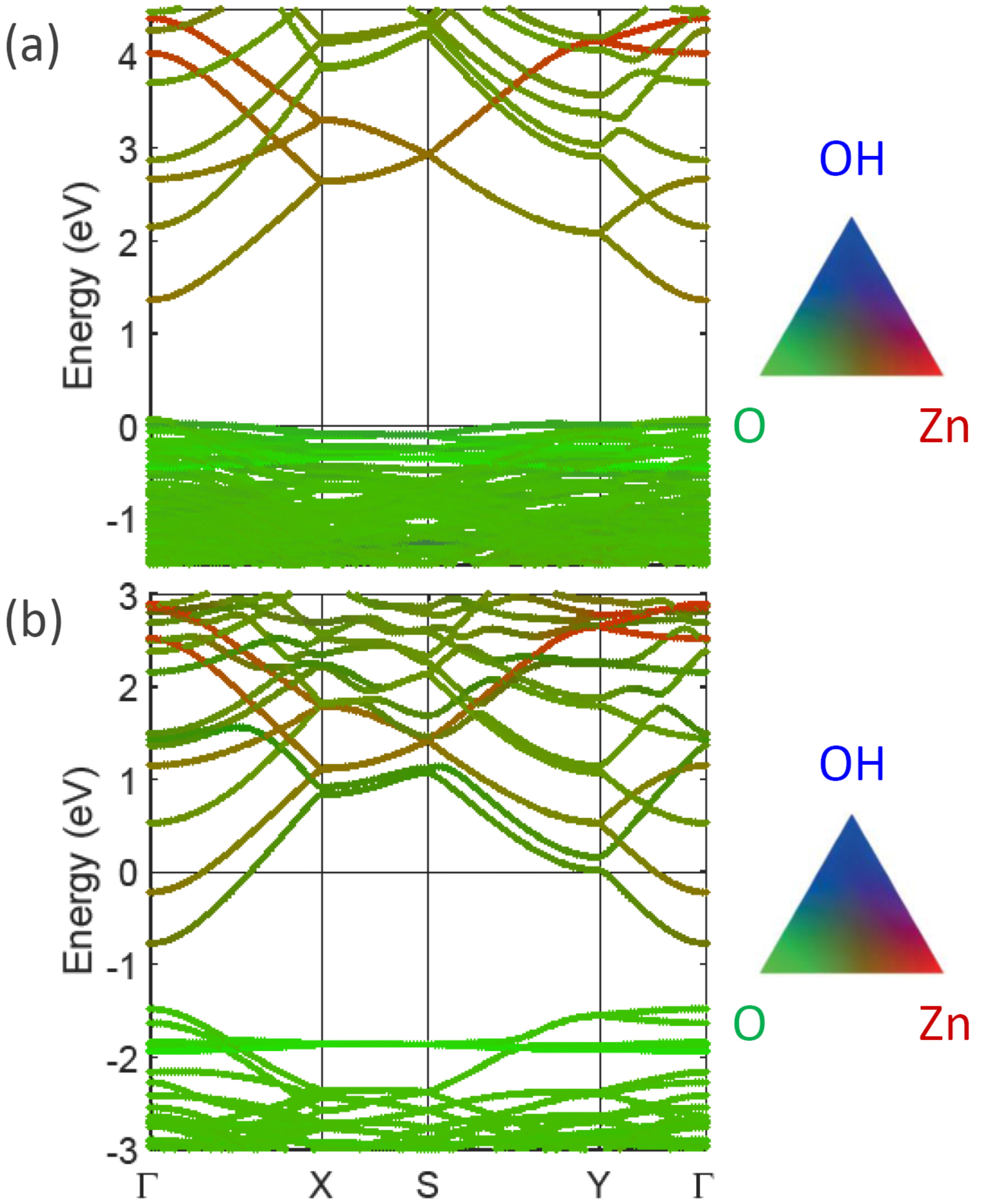}
\end{center}
\caption{\label{fig12} Character of the band structures
of (a) the undoped ZnO slab with OH adsorbed, with the colors highlighting the
OH, Zn, and O contributions (blue, red, and green, respectively), and (b) similar for the $S_2$ slab.}
\end{figure}

%figure13*
\begin{figure}
\begin{center}
\includegraphics*[width=0.8\hsize]{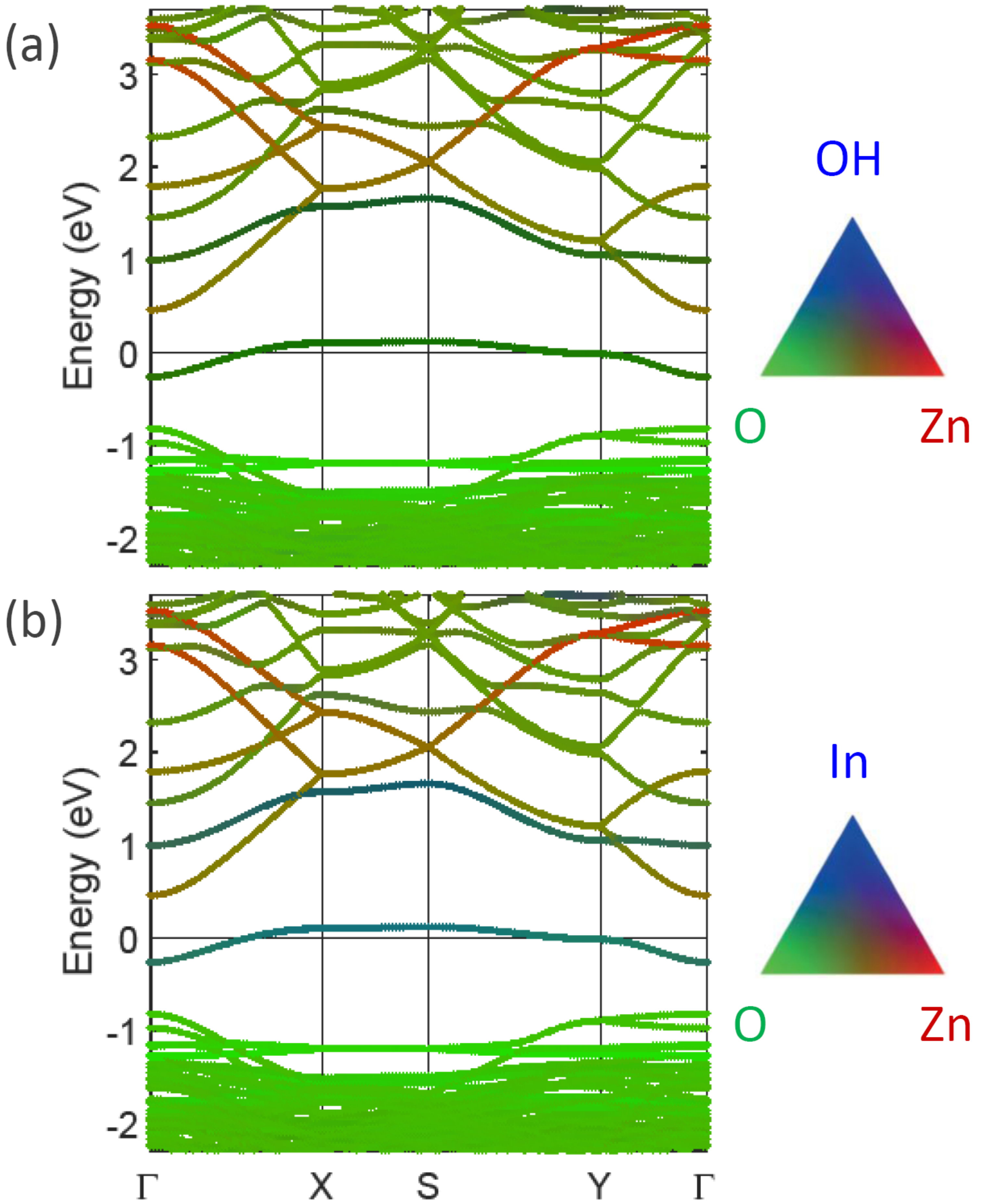}
\end{center}
\caption{\label{fig13} Character of the band structure
of the $S_1$ slab with OH adsorbed, showing in
(a) the OH, Zn, and O contributions (blue, red, and green, respectively), and in (b) the In, Zn, and O contributions,
blue indicating In in this plot.}
\end{figure}

We now consider briefly the effects of OH adsorption.
Figure~\ref{fig12}(a) shows that on an undoped
slab OH adsorption leads to empty states around the VBM.
This results from the strong hybridization of the OH levels
with Zn and O upper surface levels. This indicates that the
OH molecule will be chemisorbed, in agreement with what
was suggested by Figs.~\ref{fig7}, \ref{fig8}, and
\ref{fig9}.
Thus the surface will be $p$-type, although the mobility of
the holes will be low because of the high hole mass.
In Fig.~\ref{fig12}(b) we consider the $S_2$ slab with OH
adsorbed. The plot shows that the slab will continue to be
$n$-type, but with a noticeable charge carrier decrease
[compare with Fig.~\ref{fig10}(a)]. This is due to the
strong charge transfer involved in the OH-Zn bridge bonds
depicted in Fig.~\ref{fig6}(b) [see also Fig.~\ref{fig8}(b)].
In Fig.~\ref{fig12}(b) the OH levels are deep, below the
lowest energies shown. Indium does not contribute
significantly in the energy range in the figure, so the
corresponding plot is not shown.

Finally, in Fig.~\ref{fig13} we consider OH adsorbed on
slab $S_1$. As the plots show, OH tends to passivate the
In dangling bonds at the surface similar to CO.
This suggests that the adsorbed OH molecules
will interfere with the signals produced by CO adsorption.
Thus, overall, OH would have a negative impact on the CO
sensing properties of the In-doped slabs.

\section{Summary and conclusions}

In summary, we present a first-principles computational study of CO and OH adsorption on non-polar
ZnO (10$\bar{1}$0) surfaces doped with indium. We find that
In substitutes Zn preferentially at the surface. The presence of In atoms at the surface of ZnO favors CO adsorption, resulting in an elongation of the C-O bond and in charge transfer to the surface. Our charge transfer and
band structure analysis shows that CO tends to passivate
the In dangling bonds when it is located at the surface. This
indicates that In doping at the surface of ZnO should increase
the electronic response of the latter upon adsorption of CO. 
On the other hand, the adsorption of OH molecules on
the surfaces we studied will tend to have a negative impact on
their CO sensing properties. It is important to point out,
however, that further work is needed in order to have a more
complete picture of the phenomenology of CO adsorption on
In-doped ZnO surfaces. Indeed, the dissociative
adsorption of H$_2$O should be considered, as this is often
the source of OH groups at the surface in humid conditions.
The adsorption of H, O, and O$_2$ should also be
studied. In addition, ZnO surfaces can be expected to present
native defects, which can also play an important role when it
comes to CO adsorption.

\section*{Acknowledgments}

We acknowledge the financial support of FWO-Vlaanderen
through project G0D6515N and of the ERA.Net RUS Plus initiative
through grant No. 096 (FONSENS).
The computational resources
and services used in this work were provided by the VSC (Flemish
Supercomputer Center) and the HPC infrastructure of the University
of Antwerp (CalcUA), both funded by FWO-Vlaanderen and the
Flemish Government-department EWI.

\appendix
\section{}

\subsection{Comparison with HSE calculations}

Electronic structures obtained within a DFT$+U$ approach
depends importantly on the $U$ value used. It has a strong
influence, for instance, on the calculated band gap value of
semiconductors. Some of our main conclusions are based on
the analysis of the band structure of the slabs we studied.
It is important, thus, to validate the approach in this
study. The HSE hybrid functional offers a method of choice
for a comparison because of its reliability in determining the
electronic structure of semiconductors.
Unfortunately, it is
computationally demanding, and cannot be applied systematically
to study large systems like the present ones.
Therefore, here we focus on the band structures resulting in
two of the most crucial observations in our study, namely
the localized states in the band gap in the case of the
$S_1$ slab, and their passivation by CO adsorption.

\begin{figure}
\begin{center}
\includegraphics*[width=0.8\hsize]{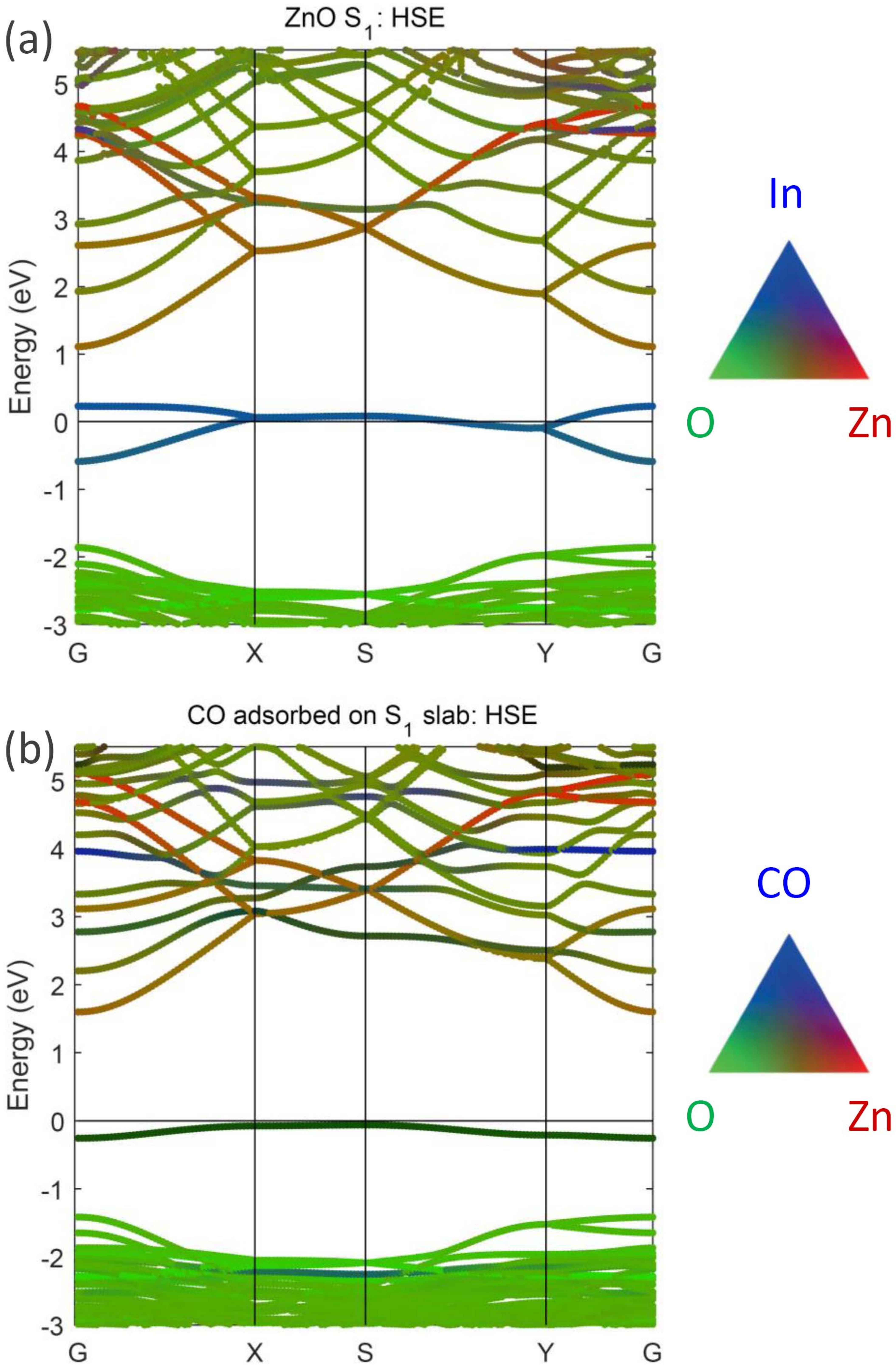}
\end{center}
\caption{\label{fig14} Character of the band structure
of the $S_1$ slab with OH adsorbed, showing in
(a) the OH, Zn, and O contributions (blue, red, and green, respectively), and in (b) the In, Zn, and O contributions,
blue indicating In in this plot.}
\end{figure}

In Fig.~\ref{fig14} we present the HSE band structures
corresponding to Figs.~\ref{fig11}(a) and (b). The HSE
calculations
were perfomed using the same atomic positions, same 
{\bf k}-point grid, and same computational parameters as in
the DFT$+U$ calculations. The HSE calculations were done using
a 0.375 fraction of exact Hartree-Fock exchange, which
correctly reproduces the experimental band gap of
wurtzite ZnO \cite{janotti11}. Fig.~\ref{fig14}(a) should be
compared with Fig.~\ref{fig11}(a),
and Fig.~\ref{fig14}(b) with Fig.~\ref{fig11}(b).
The main differences are a wider gand gap in the HSE
case, as one could expect, and band dispersions (band widths).
However, the presence and nature of the states in the band
gap in Figs.~\ref{fig11}(a) and \ref{fig14}(a) is 
essentially the same, being localized and of dominant In character, with some admixture of oxygen content.
Moreover, the passivation of these gap states by CO adsorption,
with otherwise little effect on the electronic structure,
takes place in the most similar way in both calculations,
as the comparison of Figs.~(b) shows.

The above shows that our band structure analysis in the main
text and the conclusions drawn thereupon are sound.

%figure appendix*

\subsection{CO on $S_1$: charge density difference}

\begin{figure}
\begin{center}
\includegraphics*[width=0.8\hsize]{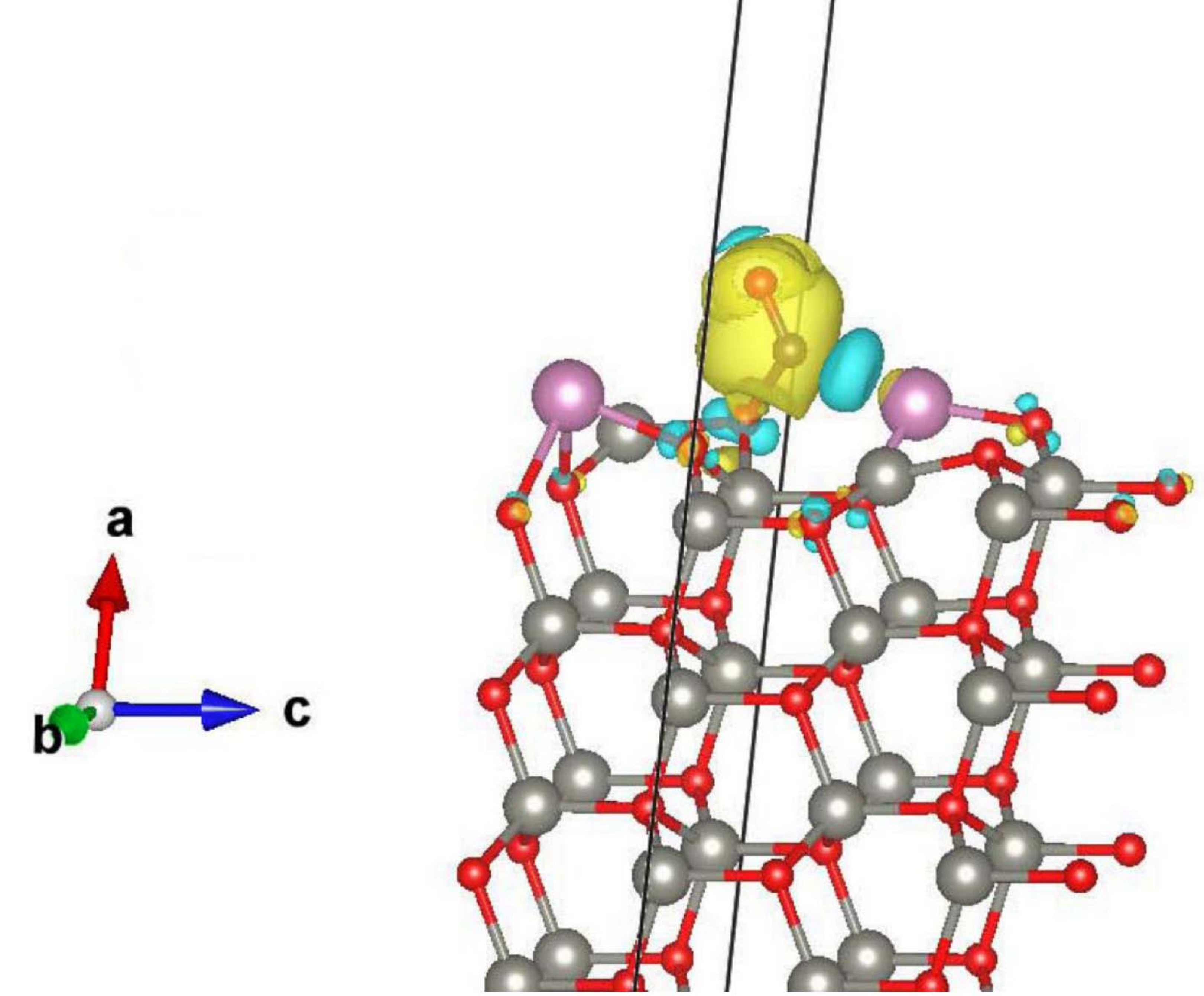}
\end{center}
\caption{\label{fig15} Plot of the charge density difference,
as defined in Eq.~(\ref{chdd}).
In Fig.~\ref{fig5}(c) one can
see that the CO molecule sits at the extreme left side of the
top surface. For clarity, in the above figure
the plotting window has been shifted so that the CO molecule
is at the center of the top surface.}
\end{figure}

In order to shed more light onto the nature of the CO
bonding to the surface in the case of slab $S_1$,
we present in Fig.~\ref{fig15} below
a charge density difference plot, defined as
\begin{equation}
\delta\rho=\rho_{{\rm CO\,on}\,S_1\,{\rm slab}}-
\rho_{S_1\,{\rm slab}}-\rho_{\rm CO\,alone}. \label{chdd}
\end{equation}
The charge density difference $\delta\rho$ varies between
$-1.63 e$ and $+0.07 e$ per unit volume. The plot shows the isosurface for
$|\delta\rho|=0.0075 e$ per unit volume, with yellow indicating a positive charge density
and cyan a negative charge density
(the isosurfaces for
larger absolute values rapidly shrink around the ions, and
information on bonding is lost). The plot can be readily
interpreted as illustrating how CO tends to passivate the
In dangling bonds, in support of our discussion on
Fig.~\ref{fig11} and one of the main conclusions of our
work.

%figure appendix*

\end{document}